# Early detection of diabetes through transfer learning-based eye (vision) screening and improvement of machine learning model performance and advanced parameter setting algorithms

Mohammad Reza Yousefi [1,2] *, Ali Bakrani [1], Amin Dehghani [3]

[1] Department of Electrical Engineering, Najafabad Branch, Islamic Azad University, Najafabad, Iran

[2] Digital Processing and Machine Vision Research Center, Najafabad Branch, Islamic Azad University, Najafabad, Iran

[3] School of Electrical and Computer Engineering, University of Tehran, Iran

* Corresponding author: Mohammad Reza Yousefi

mr_yousefi81@yahoo.com

**Abstract**

Diabetic Retinopathy (DR) is a serious and common complication of diabetes, caused by prolonged high blood sugar levels that damage the small retinal blood vessels. If left untreated, DR can progress to retinal vein occlusion and stimulate abnormal blood vessel growth, significantly increasing the risk of blindness. Traditional diabetes diagnosis methods often utilize convolutional neural networks (CNNs) to extract visual features from retinal images, followed by classification algorithms such as decision trees and k-nearest neighbors (KNN) for disease detection. However, these approaches face several challenges, including low accuracy and sensitivity, lengthy machine learning (ML) model training due to high data complexity and volume, and the use of limited datasets for testing and evaluation.

This study investigates the application of transfer learning (TL) to enhance ML model performance in DR detection. Key improvements include dimensionality reduction, optimized learning rate adjustments, and advanced parameter tuning algorithms, aimed at increasing efficiency and diagnostic accuracy. The proposed model achieved an overall accuracy of 84% on the testing dataset, outperforming prior studies. The highest class-specific accuracy reached 89%, with a maximum sensitivity of 97% and an F1-score of 92%, demonstrating strong performance in identifying DR cases. These findings suggest that TL-based DR screening is a promising approach for early diagnosis, enabling timely interventions to prevent vision loss and improve patient outcomes.

# 1. Introduction

Early detection of diabetes is of paramount importance and helps prevent serious complications, including damage to the kidneys, heart, blood vessels, and eyes [1–5]. Besides, faster diabetes treatment and careful management can assist in improving blood sugar control and reducing the likelihood of side effects. Diabetic retinopathy (DR) screening through retinal images can help prevent serious injuries, e.g., retinal damage, and subsequently, further control blood sugar levels, and facilitate eye (vision) improvement [6–10]. Additionally, early detection enables individuals to control their blood sugar levels by making changes to their diet and physical activity, and subsequently, reducing potential complications. Moreover, diabetic patients (people with diabetes) can deal with their disease more effectively by improving their quality of life and optimal management. Finally, early detection helps avoid higher treatment costs during advanced stages of the disease, enabling patients to reap the benefits of existing treatments by minimizing the need for more complicated treatments [11–14]. DR is a serious and common complication in diabetic patients, which greatly affects the retina, a part of the eye responsible for converting light into neural signals, which plays a fundamental role in vision. High blood sugar levels in diabetic patients can directly damage small retinal blood vessels, leading to retinal hemorrhage and macular edema (retinitis) disorders, and subsequently, gradual vision loss. Since diabetes usually starts without any [warning] symptoms, people should refer to specialists for periodic eye examinations because early detection and proper management can prevent serious vision problems. First and foremost, the treatment process includes accurate blood sugar control, proper diet, medications, and, if necessary, surgical treatments [15]. Fig. 1 compares the retinal differences between diabetic patients (damaged due to high blood sugar) and non-diabetic people.

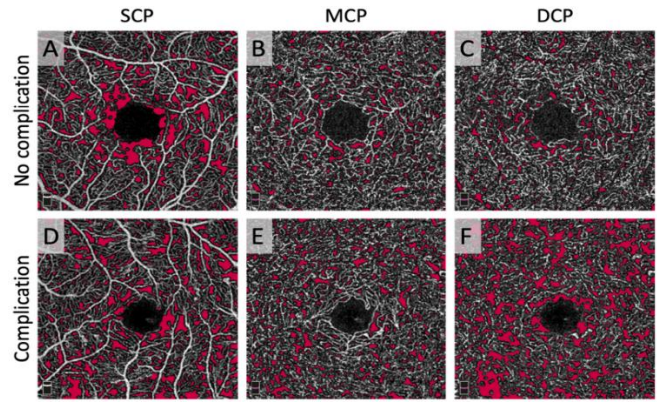

**Fig. 1:** Retinal images of DR patients vs non-diabetic people [16]

Early detection of diabetes through retinal images by collecting, pre-processing, and extracting retinal image features, along with ML algorithms, allows faster prevention and intervention to face the symptoms of diabetes. This method helps in improved early detection of retinal changes, and subsequently, providing timely treatments and interventions, and finally, preventing the development of serious vision problems caused by diabetes [17]. Neural networks are a type of ML system inspired by human brain structure, consisting of processing units called densely interconnected artificial neurons. They can undertake various tasks through the learning process, during which the network extracts patterns and rules from the input data and uses them for prediction, classification, etc. [18]. Diabetes is one of the chronic diseases worldwide with an increasing morbidity rate. In addition to general physical health, diabetes directly affects eye and vision health. DR is one of the serious complications of diabetes for eye and vision, which can result in visual impairment and blindness. Due to its high significance and direct impact on diabetic patients' quality of life, early detection of DR and diabetes through TL-based eye (vision) screening must be studied to prevent the development of diabetes-related complications. This study employs neural networks. Some advantages of neural networks include:

- Ability to learn from data: Neural networks can learn independently through data analysis and performance improvement.
- Ability to solve complex problems: Neural networks can solve complex problems, which are difficult, if not impossible, to solve by traditional methods.
- Ability to generalize to new data: Neural networks can predict and classify new data by the knowledge obtained from training data [19].

Fig. 2 illustrates the block diagram of a neural network.

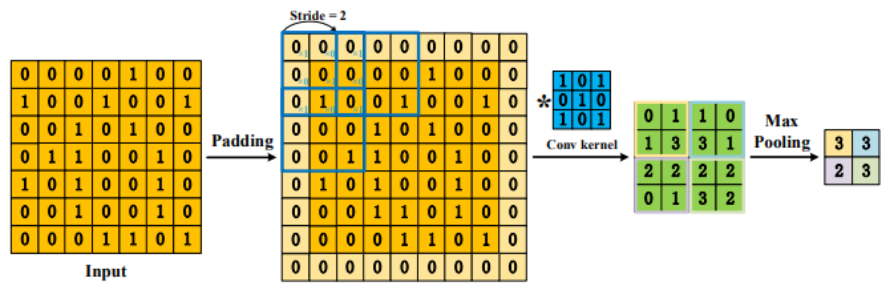

**Fig. 2:** Block diagram of a neural network [18].

This study was conducted for the early detection of diabetes through retinal image screening by neural networks with TL focused on DR (DR-focused TL). It was performed on TL and ML model performance improvement using methods such as dimensionality reduction, learning rate setting (adjustment), and advanced parameter setting algorithms, which is an essential research innovation. In dimensionality reduction, information is displayed in a smaller space usually by techniques like principal component analysis (PCA). In the learning rate setting, the ML model's learning rate is set (adjusted) to accelerate and improve the learning quality. In advanced parameter setting algorithms, the ML model parameters are optimized [20]. For this purpose, the data is divided into three sets: training (70%), validation (15%), and testing (15%).

## 2. Literature Review

In [8], different DR stages were detected from retinal images obtained from the IDRiD dataset using ConvNets, where 80% of the data was assigned to training and the remaining 20% to testing. Ensemble learning models consisted of five CNNs: Resnet50, Inceptionv3, Xception, Dense121, and Dense169. According to the results, the above models succeeded in diagnosing the disease with higher accuracy than the previous methods, especially in the early stages.

In [21], an automated DR detection model was proposed using retinal images collected from Melaka Hospital. This model was trained by three neural networks: BNN, DNN, and CNN. According to the results, CNN outperformed NN and DNN with an accuracy of 93%. Furthermore, target class thresholds were identified using a weighted fuzzy C-means (FCM) algorithm. A new segmentation method, known as EAD-Net, was proposed for feature extraction and pixel-wise label prediction using a ConvNet (sensitivity, specificity, and accuracy of 97% in the *e_ophtha_EX* dataset) [22]. Due to their simpler structure, conventional ANNs generally possess lower processing time and accuracy than their deep counterparts. Image processing is more commonly performed by deep CNNs. In [23], the "DeepDR" system was developed by taking (capturing) 466,247 retinal images from 121,342 diabetic patients for the detection of early to later DR stages. In the above system, the area under the curve (AUC) of the performance characteristic was calculated to be 0.901, 0.941, 0.954, and 0.967 for detecting microaneurysms, cotton wool spots, and hemorrhages, respectively. Also, these figures were 0.943, 0.955, 0.960, and 0.972, respectively, for disease classification.

In [24], a clinical decision support system (CDSS) (medical decision support system (MDSS)) was proposed, which comprises four basic steps: imaging, image preprocessing (including localization of retinal structures), feature extraction, and

DR classification. A combination of blurred image processing, Hough circle transform, and feature extraction techniques were adopted to improve the performance. The relevant database is located in Kaggle with 58% accuracy and 85% accuracy [24].

An attention module, called CBAM, was proposed to infer the attention map along two dimensions: channel and spatial. These attention maps are then multiplied by the input feature map for (adaptive) feature optimization. CBAM is a lightweight and general module that can be integrated into any CNN architecture. It was tested on three datasets, namely ImageNet-1K, MS COCO, and VOC 2007 (the highest accuracy: 78%) [25].

In [26], DR detection was carried out using CNN-based DL algorithms that classify images into two classes. A model with Siamese-like architecture was trained using the TL technique, which accepts binocular (fundus) images as inputs to learn their correlation. The results indicated the positive effect of binocular design on model efficiency and their superiority over existing monocular models.

In [27], CNNs (e.g., AlexNet) were adopted to classify retinal images of the Messidor dataset based on disease severity. This algorithm achieved an accuracy of 96% for healthy and unhealthy retina images in steps 1, 2, and 3.

In [28], a modified U-Net architecture was proposed for the segmentation of lesions from images to assess disease severity, which defines the targeted region through periodic shuffling and sub-pixel convolution. This model was trained on two publicly available datasets, namely IDRiD and e-ophtha, which obtained a similar accuracy of 99%.

In another study, a deep CNN method was proposed to segment four retinal lesions, which utilizes a collaborative attention mechanism (CAM) architecture and classification to reduce model overfitting. According to the results of the tests performed on the Fundus dataset, the average ACU was calculated to be 67% [29].

In [30], a relation transformer block (RTB) was proposed to segment retinal lesions (for retinal lesion segmentation). This model uses self-attention and mutual attention transformers to exploit global dependencies among lesion features and vascular information. To this end, it initially takes small lesion patterns and stores the detailed information in a deep network. The IDRiD dataset was utilized, and the proposed method achieved an accuracy of 94%.

Given the considerable importance of the early detection of DR and its positive effect on the general health and quality of life of diabetic patients, it is quite necessary and justified to conduct research in this field. Studies can help improve individual health, reduce DR treatment costs, and prevent dangerous complications for diabetic patients.

In [31], a suitable DR severity grading framework was proposed. The advantages of ML and DL algorithms were pooled to establish a strong framework through a trade-off between model processing and classification performance. For this purpose, the fundus images (FIs) were initially preprocessed using CLAHE to highlight the lesions more clearly, then normalized and finally reshaped. A lightweight deep CNN model was then developed to extract the most discriminant features from the processed FIs. The extracted features were standardized to be imported into the extreme learning machine (ELM) algorithm for DR severity level classification. The model proposed in [31] produced promising results for 34,984 data (dataset 1) and 3,662 FIs (APTOS, 2019). Not only did it show higher classification performance but it also significantly reduced parameters, layers, and processing time. This model outperformed existing state-of-the-art (SOTA) models for both datasets and succeeded in the early detection of DR severity with good accuracy. Hence, it helped reduce patients' vision loss and spare doctors' precious time.

In [32], a robust and efficient model was developed for automated DR detection (diagnosis) focused on extracting mostly descriptive and distinctive deep features, and subsequently improving DR detection performance. To get an optimal representation, features were extracted from multiple pre-trained CNN models and blended using pooling approaches (multi-modal pooling module). These final representations were used to train a deep neural network (DNN) with dropout at the input layer. The proposed model comprises three distinct modules, namely feature extraction, model training, and evaluation (assessment) module, where the initial representation of retinal images was obtained from pre-trained VGG16, NASNet, Xception Net, and Inception ResNetV2. Since each pre-trained model was waiting for different-sized input images, the given retinal images were reshaped based on the input dimensions accepted by them. For example, the images were reshaped to 224 x 224 x 3 when VGG16 was used. Fig. 3 visualizes the feature maps resulting from the final convolution blocks of the VGG16 and Xception models when sending retinal images.

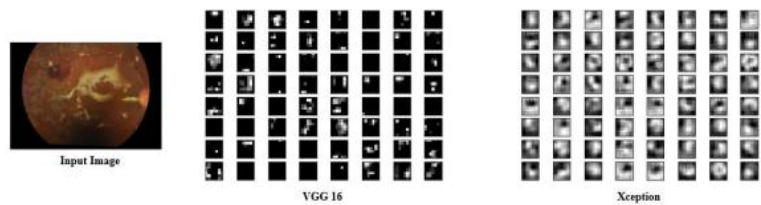

**Fig. 3:** A representation of the final convolution blocks of VGG16 and Xception models in retinal images as inputs [32].

In [32], two distinct pooling approaches (one-dimensional pooling and cross-pooling) were proposed to combine deep multimodal features extracted from VGG16 (fc1 and fc2) and Xception. While 1D pooling is utilized to select locally salient features from each VGG16 region, cross-pooling allows the pooling of salient features obtained by 1D pooling through a global representation of Xception. Each feature element *ui* from the output vector *U* is calculated by one of the following three methods, as in Eqs. (1)-(4):

$$1D\ Max\ pooling: \hat{u}_i = max(u_{i*2}, u_{i*2+1})\ \forall i \in \{1, 2...d_2\} \quad (1)$$

$$1D\ Min\ pooling: \hat{u}_i = min(u_{i*2}, u_{i*2+1})\ \forall i \in \{1, 2...d_2\} \quad (2)$$

$$1D\ Averagr\ pooling: \hat{u}_i = meam(u_{i*2}, u_{i*2+1})\ \forall i \in \{1, 2...d_2\} \quad (3)$$

$$1D\ Sum\ pooling: \hat{u}_i = u_{i*2}, u_{i*2+1};\ \forall i \in \{1, 2...d_2\} \quad (4)$$

In [33], the authors employed 3,662 images from the Kaggle dataset (70% for training and 30% for testing). To prepare the images for the model, the images were first resized to 224x224x3.

The proposed method should be adopted as an important training method in image enhancement by horizontal and vertical flip, zoom, and channel shift features. The dataset test used batch size 32 and binary crossover as the loss function, ReLU as the activation function, and sigmoid as the activation function in the final layer of the model to reduce model overfitting and increase its accuracy. Adam was also used as an optimizer, and the images were classified into five classes (accuracy = 71.18%).

In [34], the authors focused on the accurate detection of DR and diabetic macular edema (DME) based on retinal FIs. Disease severity was determined based on the features extracted from the images. Fig. 4 depicts an architectural view of the proposed work. The main idea behind OPTICS (Ordering Points to Identify the Clustering Structure) is to extract the clustering structure of a dataset by identifying the density-connected points. This approach creates a density-based representation of the data by creating an ordered list of points called a reachability plot. Each point in the list is associated with a reachability distance, which is a measure of how easy it is to reach that point from other points in the dataset. Therefore, points with similar reachability distances are likely to be in the same cluster.

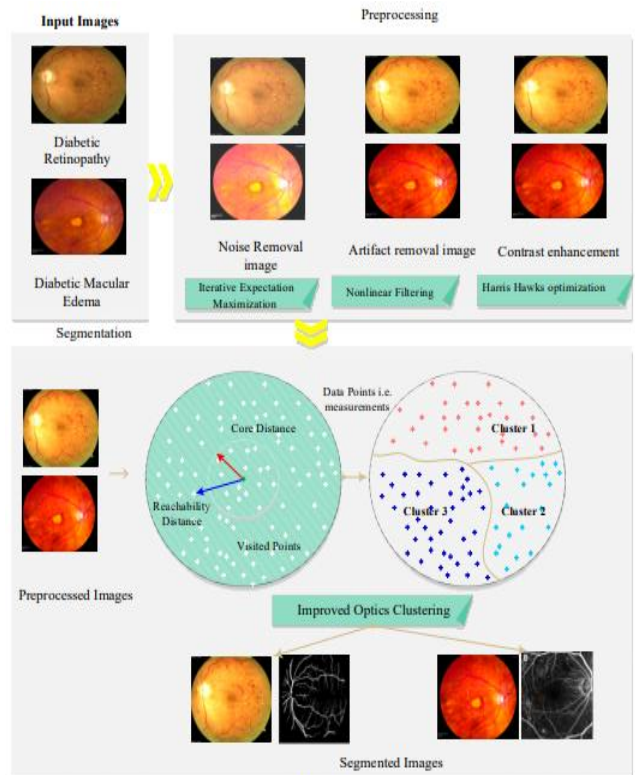

**Fig. 4:** The proposed model in [34].

In [35], a general DR architecture classification method was presented (Fig. 5). The DR model is composed of two stages: 1) training and validation and 2) testing. In the first stage, the dataset was collected from multiple sources, including hospitals and various paid data repositories. The private dataset consists of 57,625 DR images. The experts labeled the dataset as either DR-positive or DR-negative. This dataset was further divided into two subsets: training (80%) and validation (20%). After splitting the dataset, the subsets served as inputs for the CNN-based DR model to train and validate the model. The proposed DR model consists of convolutional and pooling layers.

In convolutional layers, convolutional operations with different filter sizes were performed to extract features from the input dataset. Moreover, the Rectified Linear Unit (ReLU) activation function (AF) was applied to convert linear data to nonlinear data. The pooling layer was activated after the convolutional layer was used. Common types of pooling layers include max pooling, min pooling, average pooling, and sum pooling. Herein, the pooling layer was applied to reduce the size of the activation map without losing important information in the dataset.

Subsequently, the convolutional and pooling layers and the feature map are converted into a vector. The DR model employs multiple fully connected layers (FCLs). Finally, the output was obtained from the DR model in the form of DR-positive and DR-negative results. In the second phase of the study, real-time DR data in the testing dataset were acquired from the patients at the Sindh Institute of Ophthalmology & Visual Sciences (SIOVS), Hyderabad, Pakistan, using image-capturing devices. A real-time testing dataset consisting of images obtained from patients with Type II diabetes (T2D) was created for five weeks. An intelligent model was then used to evaluate the quality of real-time test dataset images. Low-quality images detected by the intelligent model were rejected and then recaptured [35].

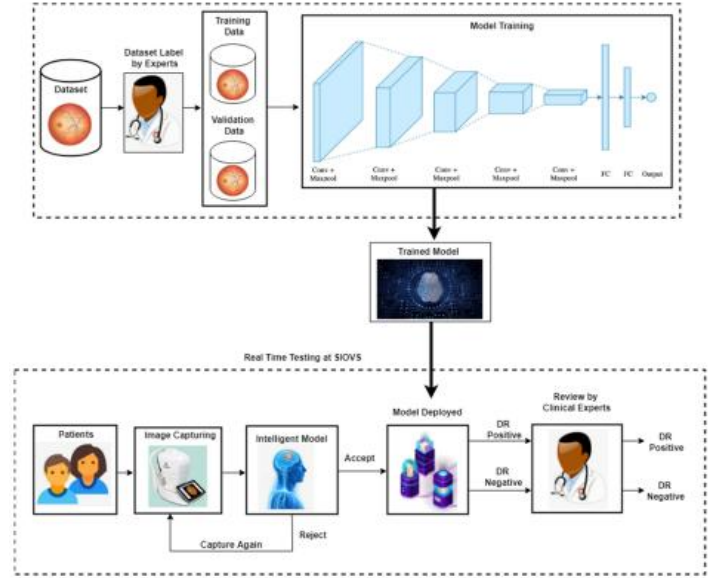

**Fig. 5:** DR detection architecture [35].

In [24], the number of classes was reduced from 5 to 3 for multiclass classification, namely NDR (stage 0), MDR (stage 1-2), and PDR (stage 3-4). The lowest number of images was determined in three classes, and the same number of images were randomly selected from other classes. This method yielded the fewest images (n=488) in the combination of stage 3–4 labeled class (PDR). Therefore, 488 images were randomly selected from the remaining classes (i.e., NDR and MDR). These images were transferred to CNN models to extract feature vectors, and the batch size was set to 32. The feature vectors obtained from each model were combined to form a hybrid feature vector to transfer to different classifiers. For further comparison, individual TL models, which only use DenseNet or ResNet feature vector of 1000 features, were also transferred to the classifiers. The hybrid model proposed in [24] achieved the highest accuracy of 97.8% for binary classification and 89.29% for multiclass classification using the support vector machine (SVM) classifier, indicating its superiority over recent similar binary and multiclass DR detection approaches.

## 3. Materials and Methods

### 3.1. Proposed Method and Data Utilization

This study hypothesizes that TL, advanced parameter setting, and image dimensionality reduction techniques can significantly improve ML model performance in novel applications. Image dimensionality reduction is expected to reduce model training time while preserving essential image features. Optimizing the learning rate is anticipated to positively influence model performance when the learning rate reduction function is properly tuned. Parameter setting algorithms are proposed to enhance model stability while minimizing performance fluctuations.

To evaluate these aspects, the study utilizes datasets that include retinal images of diabetic patients (n=35,126; 28x28-pixel color images) for diabetic retinopathy detection and grading.

### 3.2. Datasets

The data utilized here (supervised and labeled) were extracted from the Kaggle database (https://www.kaggle.com/datasets/tanlikesmath/diabetic-retinopathy-resized/data). This research was simulated in the Python programming environment. Furthermore, a certain function was implemented to avoid randomized results (using corresponding functions) in parallel with CUDA and graphics processing modules in PyTorch.

### 3.3. Data Preprocessing

This section is dedicated to image data preprocessing, cleaning (cleansing), and preparation. The preprocessing technique is adopted primarily to eliminate unwanted noise and optimize certain image features. Therefore, the following steps are done:

As can be seen in the above piece of code:

1) '*image_size = 256*' indicates optimal image dimensions; that is, the images are all resized to the above dimensions.

2) The '*prepare_image*' function is employed for image preprocessing. Here, input parameters include '*path*' (image path), '*sigmaX*' (image sharpening control parameter), and '*do_random_crop*' (do or failure to do random crop).

### 3.4. Image Cropping

First, it is determined whether an image is one- or three-channel. If the image is one-channel, a mask is generated from values greater than the threshold, and the corresponding image is extracted. On the contrary, if the image is three-channel, it will initially be converted to black and white. Then, the mask is regenerated, and the masked image is extracted from the original image based on the mask coordinates. Finally, the output image of this operation is returned. Note that this function is used to remove the black parts of the image and increase their resolution.

### 3.5. Training and Testing Data

Training datasets are used to train the model, during which validation datasets evaluate model performance. Of the total data in the dataset used herein, 28,000 data were assigned to training and the rest to testing.

### 3.6. Histogram representation of class label distribution

Fig. 6 displays the process of creating a histogram with proper axes called *Class Distribution* by executing a piece of code written below, indicating the number of samples belonging to each label (eye resolution level).

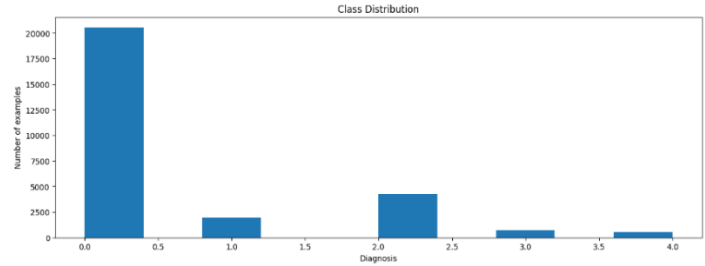

**Fig. 6:** Representation of class label distribution (the number of samples belonging to each label)

## 4. Results and Discussion

### 4.1. Retinal image representation

This section demonstrates several sample images from the training dataset using different transformations. In computer science, transformation refers to converting data from one format to another. Fig. 7 clearly depicts the structural and functional changes in cells and tissues, such as DR in T2D, as well as the transformations applied to the data. This step ensures transformation accuracy (precision) and image preparation in a proper format to be imported into the model.

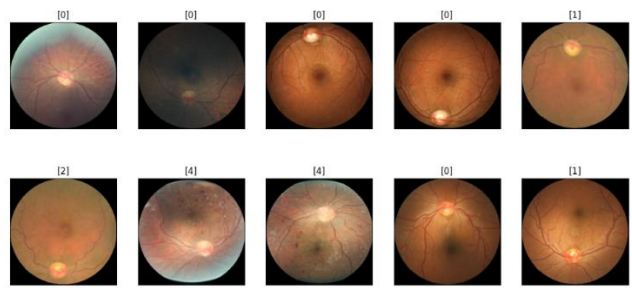

**Fig. 7:** Training dataset using different transformations

It should be noted that each image is read from the data based on its path and name (title). The calculated information includes '*height*', '*width*', '*number of channels*', and '*ratio*' (width-to-height ratio).

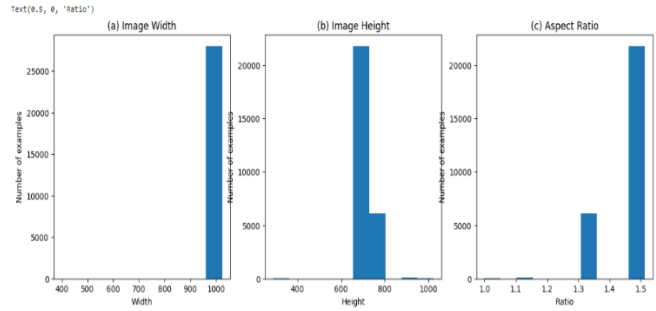

**Fig. 8:** Converting the list of calculated information to a *DataFrame*

Columns are named like ['*level*', '*height*', '*width*', '*channels*', '*ratio*'], respectively. The above information can be exploited to analyze and better understand the image characteristics as well as to apply some changes, such as image resizing. Various transformations were defined for training, validation, and testing data, which are utilized to increase data diversity and improve model performance during training.

### 4.2. Improvement of the proposed performance model

As previously mentioned, transformations were applied to increase data diversity and improve model performance during training. This step was taken to ensure the accuracy (precision) of transformations and image representation according to Fig. 9.

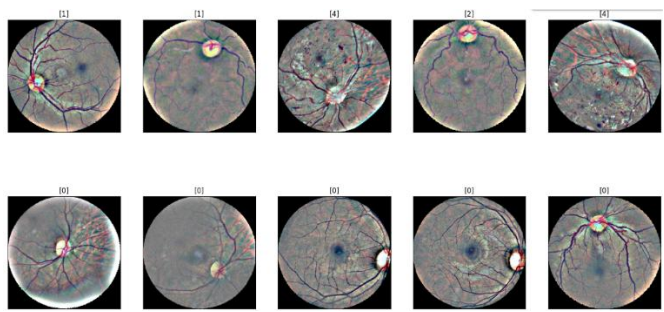

**Fig. 9:** Improvement of model quality in training transformation representation

Since testing data indicators are readjusted, an '*EyeData*' class sample is defined, including testing data (10 initial samples), image path, and testing (data) transformations. A data loader is then developed for the test data. Afterward, a batch of data is read. Finally, the data loader is applied to ensure the accuracy (precision) of transformations and image representation during testing according to Fig. 10.

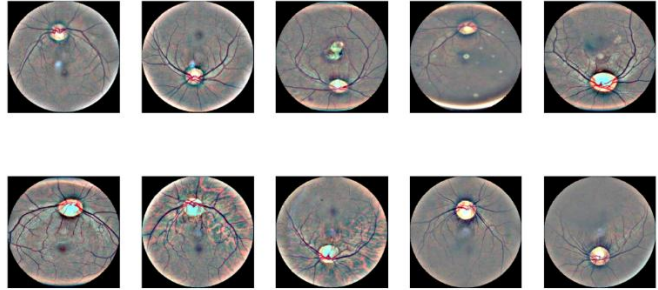

**Fig. 10:** Image representation during testing

### 4.3. Saving validation results in each epoch

In the training phase, the model proposed herein is put into **(model.train())** mode. Then, the inputs and labels are transferred to the GPU. The model then executes the forward- and back-prop processes, updates the weights, and calculates and saves the training error. In the validation phase, the proposed model is put into **(model.eval())** mode and the inputs and labels are taken from the cluster.

As in the testing phase, the inputs and labels are transferred to the GPU. Model predictions on the validation data are calculated and saved in **fold_preds**. Finally, the validation error is calculated and saved. Fig. 11 shows the code output as the results of this section as a graph.

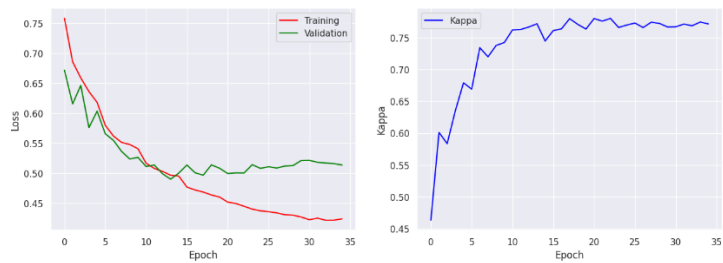

**Fig. 11:** Error dynamics and Kappa

### 4.3.1. Error Plot

- The diagram on the left (Fig. 11) shows error dynamics during the training and validation periods.

- The red and green lines represent training and validation errors, respectively.

- The horizontal and vertical axes indicate the number of epochs and the error rate, respectively.

### 4.3.2. Kappa Plot

- The diagram on the right (Fig. 11) shows Kappa dynamics during training.

- The blue line represents the Kappa criterion on the validation data.

- The horizontal and vertical axes indicate the number of epochs and the Kappa criterion, respectively.

## 4.4. Evaluation of the final model performance

This piece of code evaluates the final model performance on validation data, including calculating out-of-fold (OOF) error and Kappa. According to the results, the OOF error and Kappa are approximately 0.5004 and 0.7801 on the validation data for the proposed model, indicating its good performance and consistency with the validation data. Here, a Kappa criterion of 0.6-0.8 indicates good performance and a Kappa criterion of 0.8-1 indicates excellent performance. Besides, the OOF error should be minimal and close to 0, which is acceptable here.

## 4.5. Confusion Matrix

In this piece of code, a confusion matrix is formed and displayed graphically. A confusion matrix is an evaluation tool to determine the number of correctly and incorrectly detected samples in each class. In Fig. 12, the horizontal and vertical axes indicate model predictions and actual labels, respectively. Also, the figures listed in each plot cell indicate the ratio (proportion) given to the number of relevant samples.

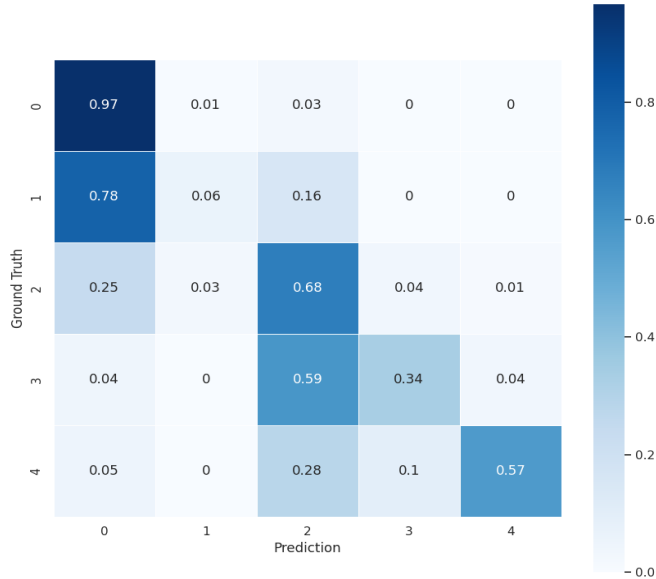

**Fig. 12:** Confusion matrix

## 4.6. Report on the proposed model performance

This section provides a detailed report of model performance on testing data, including information such as accuracy, sensitivity (Recall), F1-Score, and other evaluation criteria. Table 1 lists the above data for each class separately.

**Table 1.** Performance evaluation results of the model proposed herein on testing data

| Criteria | Accuracy | Sensitivity | F1-Score | Other criteria |
|---|---|---|---|---|
| 0 | 0.89 | 0.97 | 0.92 | 5282 |
| 1 | 0.35 | 0.06 | 0.11 | 488 |
| 2 | 0.66 | 0.68 | 0.67 | 1036 |
| 3 | 0.52 | 0.34 | 0.41 | 167 |
| 4 | 0.76 | 0.57 | 0.65 | 153 |
| Accuracy (Precision) |  |  | 0.84 | 7126 |
| Macro avg | 0.64 | 0.52 | 0.55 | 7126 |
| Weighted avg | 0.81 | 0.84 | 0.81 | 712 |

## 5. Conclusion

This study discussed TL and ML model performance improvement using methods such as dimensionality reduction, learning rate setting, and advanced parameter setting algorithms for eye damage in diabetes or DR. Diabetes is one of the most common metabolic disorders, which includes several types, with T1D and T2D being the most common. All people, both men and women (male and female), are exposed to diabetes at any age. Diabetic patients can hope for longer life expectancy and live a normal life like other people if they practice a healthy lifestyle and control their blood sugar. On the other hand, if these individuals, especially those with T2D, fail to control their blood sugar levels, their retina is exposed to damage and even blindness, known as DR in medical science.

This study was conducted for the early detection of DR complications by combining the DL technique and eye images as the main tools. TL algorithms are adopted to analyze retinal images and detect early DR symptoms (signs). This research yielded better results than other relevant studies. Reportedly, the proposed model achieved an accuracy of about 84% on training (testing) datasets. By providing detailed information on each class, the above report can provide you with deep insights into model performance in each classification.

The following critical points can be extracted from the above detailed report:

The accuracy criterion for each class indicates the ratio of the number of correctly detected samples to the total number of detected samples in retinal images. The maximum accuracy value obtained herein was 0.89%. The sensitivity criterion represents the ratio of the number of correctly detected samples to the total number of samples that should be detected. The maximum sensitivity value obtained herein was 0.97%, suggesting that this research yielded better results than previous studies on early detection of DR. F1-Score (accuracy + sensitivity) was another criterion reviewed herein, calculated as the geometric mean of the above two criteria. The proposed model achieved an F1-Score of 0.92%, indicating significant improvement over previous research.